# INTERMITTENT SHOT NOISE GENERATING 1/f FLUCTUATIONS

FERDINAND GRÜNEIS

*Institute for Applied Stochastic*
*Rudolf von Scholtz Straße 4, 94036 Passau, Germany*
*Email: ferdinand.grueneis@t-online.de*

When the rate of shot noise is controlled by on-off states we speak of intermittent shot noise. The on-off states lead to alternately occurring clusters of events and intermissions, respectively. We derive the power spectrum of the intermittent shot noise by applying the Wiener-Khinchin theorem. Besides reduced shot noise, we obtain excess noise, which depends on the parameters of the on-off states. We calculate the excess noise for power-law distributed on-states; within the scaling region, the excess noise is excellently approximated by $C/f^b$. The behavior of the slope $b$ and of the amplitude $C$ in dependence of the on-off times is investigated. For large scaling regions we find a preference for a pure 1/f shape. Finally, we regard the variance of events occurring within a time interval. In the presence of 1/f fluctuations, the variance of counts attains extreme values which are accompanied by an extreme property of slope $b$.

## 1. Introduction

Shot noise is a form of noise arising from the interaction of many individual and randomly occurring single events. For example, the applause following a recital in a concert hall is considered noise; it is caused by many randomly occurring claps which are summed up in our ears. A similar listening experience is provided by lead shot pellets falling onto a glass plate, which explains the origin of the term "shot noise". The random succession of elementary events is generally referred to as shot noise [1-4].

The concept of shot noise was introduced by Schottky [5] who studied current fluctuations in vacuum tubes. Shot noise arises because the electric current consists of a vast number of discrete electric charges. The current flow is not continuous but results from the motion of independent charge carriers. Shot noise is always associated with direct current flow.

When measuring the noise spectrum predicted by Schottky, Johnson [6] found an unexpected noise component at low frequencies, which is denoted flicker noise or 1/f noise. 1/f noise has subsequently been observed not only in a wide variety of electronic materials, but also in other systems, like heart rate, neuronal activity in the brain, the stock market, etc. [7]. The ubiquity of flicker noise suggests a common underlying mechanism. Despite considerable progress, researchers have not been able to agree on a unified explanation for 1/f noise. Consequently, there exist several explanations for 1/f noise.

In vacuum tubes, mobile defects (like foreign atoms or vacancies) migrating on the surface of the cathode are thought to have an impact on the work function and the current flow. This model proposed by Schottky [8] does not provide a 1/f spectrum but is appropriate to describe intermittent shot noise. The current flow is modeled by shot noise and the impact of mobile defects by a gating function. Supposed the current flow is totally blocked this leads to on-off states; this phenomenon is also called on-off intermittency.

A further example for on-off intermittency are quantum dots and other nanoparticles exhibiting fluorescence intermittency [9-10]; despite continuous excitation, the light emission of these materials switches randomly between bright (= on) and dark (= off) states. Most surprisingly, these on- and off-states follow power-law statistics. The underlying mechanism responsible for fluorescence intermittency is still a mystery. The spectrum of this two-state process shows a $1/f^b$ shape with two different slopes which can be attributed to the power-law statistics of the respective on- and off-states. The intermittent shot noise is appropriate for modelling fluorescence intermittency providing a relation between the exponents in the time- and frequency domain [11].

From a mathematical point of view, intermittent shot noise belongs to the class of doubly stochastic processes [12-14]. Thereby, the rate of a primary Poisson process is controlled by a secondary stochastic process which can be a continuous [15-17] or – like in our case – a non-continuous process.

The power spectrum of the intermittent shot noise is derived by applying the Wiener-Khinchin theorem [2-4]. The power spectrum has already been published in context to mobile defects in a semiconductor material [18], but only the results have been reported. Detailed derivations are provided in this paper. As a result, we obtain reduced shot noise and excess noise depending on the parameters of the secondary process. Special attention is paid to the case in which the excess noise exhibits a $C/f^b$ shape. We determine the amplitude $C$ and the slope $b$ as a function of the parameters of the secondary process. For large scaling regions, a preference for a pure 1/f shape is found. Finally, we investigate the variance of counts occurring within a time interval. In the presence of 1/f fluctuations, the variance of counts attains extreme values which are accompanied by an extreme property of slope $b$. The outstanding origin of 1/f noise in different systems justifies further investigations.

## 2. Shot Noise

Figure 1.a shows a time series $y_{shot}(t)$ of shot noise. The underlying point process is a Poisson process (Fig. 1.b). It is characterized by exponentially distributed inter-event times $\lambda$; the mean rate of events (= mean number of events per unit time) is $1/\lambda$. Each spike triggers an elementary pulse $h(t)$ with Fourier transform

$$H(f) = \int_{-\infty}^{\infty} h(t) \exp(-i2\pi f t)\, dt. \tag{2.1}$$

As an example, regard rectangular pulses $h(t)$ with amplitude $A$ (Fig. 1.a) leading to

$$H(f) = A \frac{1-\exp(-i2\pi f \tau_h)}{i2\pi f} = A \frac{\sin(\pi f \tau_h)}{\pi f} \exp(-i\pi f \tau_h) \tag{2.2}$$

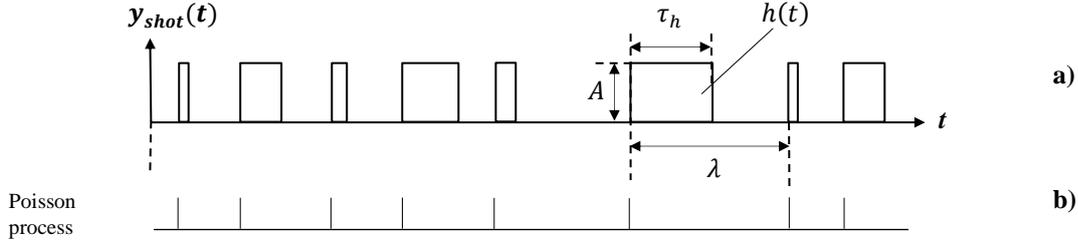

Fig. 1.a) The time series $y_{shot}(t)$ of shot noise. b) The underlying point process is a Poisson process which is characterized by independently and identically distributed inter-event times $\lambda$. Each spike triggers an elementary pulse $h(t)$ with amplitude $A$ and lifetime $\tau_h$.

For exponentially distributed lifetimes, the probability density function of $\tau_h$ is given by

$$p_{\tau_h}(t) = \frac{1}{\tau_h} \exp\left(-\frac{t}{\tau_h}\right) \tag{2.3}$$

with $\tau_h$ being the mean lifetime[1]. Using (2.2), we find after some manipulations

$$\overline{|H(f)|^2} = 2\left|\overline{H(f)}\right|^2 = \frac{2\,(A\tau_h)^2}{1+(2\pi f \tau_h)^2}. \tag{2.4}$$

Expectation values are indicated by a horizontal bar. The first term is the square value of $H(f)$, the second term the mean value of $H(f)$ squared. Applying Carson's theorem [19], the one-sided power spectrum of shot noise is given by

$$S_{shot}(f) = \frac{2}{\lambda}\overline{|H(f)|^2}. \tag{2.5}$$

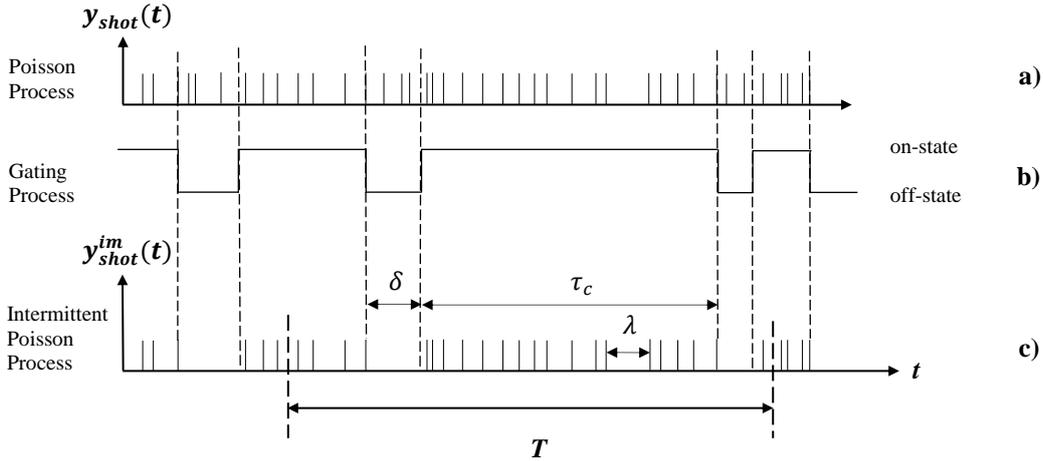

Fig. 2. a) Illustration of a Poisson point process; each spike triggers a pulse $h(t)$ as is seen in Fig. 1 leading to shot noise $y_{shot}(t)$. b) The Poisson process is gated by a two-state process with off-state $\delta$ (= intermission) and on-state $\tau_c$ (= lifetime of a cluster). c) The Intermittent Poisson Process is characterized by intermissions followed by fluctuating clusters. Each spike triggers a pulse $h(t)$ leading to intermittent shot noise $y_{shot}^{im}(t)$. The intermission starts after the last spike of the preceding cluster. This figure shows the time interval $T$ comprising three clusters; in our notation (see text) the number of clusters within $T$ is $N_c^T = 3$. The first and the last cluster are truncated by the start and the end of the time interval $T$ respectively.

---

[1] For avoiding an extended mathematical formalism, we do not distinguish between a statistical variable and its mean value.

## 3. Intermittent Shot Noise

Figure 2.a shows a Poisson Process which is intermitted by a gating process (Fig. 2.b). This phenomenon is called on-off intermittency leading to a so-called Intermittent Poisson Process (Fig. 2.c). Each spike triggers a pulse $h(t)$. The resulting stochastic process is called intermittent shot noise.

### 3.1. *The time series of the intermittent shot noise*

Figure 2.c shows the time series of the Intermittent Poisson Process. Each spike triggers a pulse $h(t)$ leading to the time series of intermittent shot noise $y_{shot}^{im}(t)$. We regard exactly $N_c^T$ clusters occurring within the time interval $T$; this is described by

$$y_{shot}^{im}(t, N_c^T) = \sum_{j=1}^{N_c^T} \eta_j(t - \Theta_j) \tag{3.1}$$

where

$$\eta_j(t) = \sum_{\mu=1}^{N_j} h_{j,\mu}(t - \vartheta_{j,\mu}) \tag{3.2}$$

is the j$^{th}$ cluster containing $N_j$ events; $\vartheta_{j,\mu}$ is the µ$^{th}$ event and $h_{j,\mu}$ the µ$^{th}$ pulse in the j$^{th}$ cluster (Fig. 3). The Fourier transform of (3.1) is

$$Y(f, T) = \int_0^T y_{shot}^{im}(t, N_c^T) exp(-i2\pi f t)\, dt. \tag{3.3}$$

Substituting herein (3.1) results in

$$Y(f, T) = \sum_{j=1}^{N_c^T} \sum_{\mu=1}^{N_j} H_{j,\mu}(f) exp(-i2\pi f \Theta_{j,\mu}) \tag{3.4}$$

with

$$\Theta_{j,\mu} = \Theta_j + \vartheta_{j,\mu} \tag{3.5}$$

being the occurrence time of the µ$^{th}$ event in the j$^{th}$ cluster (Fig. 3).

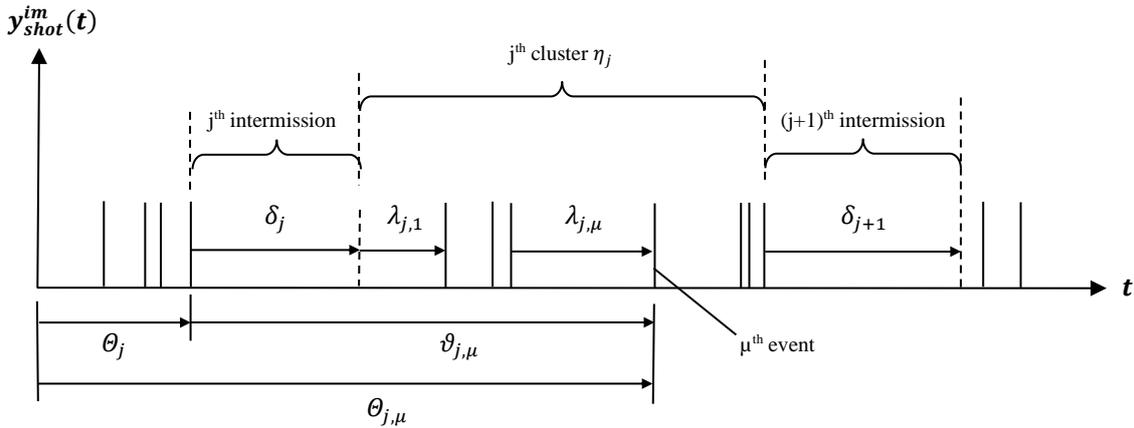

Fig. 3. Spike train of the intermittent shot noise; $\Theta_{j,\mu}$ is the occurrence time of the µ$^{th}$ event in the j$^{th}$ cluster. Each spike triggers a pulse $h(t)$ leading to intermittent shot noise $y_{shot}^{im}(t)$ (not illustrated in this Figure).

### 3.2. *Statistical features of intermittent shot noise*

The inter-event times $\lambda$, intermissions $\delta$, and pulses $h(t)$ are independent statistical variables. Fluctuating clusters are described by a cluster-size distribution

$$n = 1, 2, \dots N_{max}: \qquad\qquad q_n \tag{3.6}$$

with $N_{max}$ being a maximum cluster size. $q_n$ is the probability of finding exactly $n$ events in a cluster. The first and second moments of cluster size are denoted by

$$\overline{N_c} = \sum_n q_n\, n \qquad \text{and} \qquad \overline{N_c^2} = \sum_n q_n\, n^2. \tag{3.7}$$

Table 1 summarizes the statistical features of the Intermittent Poisson Process. The characteristic function of a statistical variable $X$ is defined by [2, 4]

$$U_X \equiv U_X(f) = \overline{exp(i2\pi f X)}. \tag{3.8}$$

The clusters can be regarded as so-called Poisson fragments with exponentially distributed inter-event times $\lambda$. The intermission $\delta$ is assumed to be exponentially distributed. For convenience we introduce the normalized off-time by

$$r = \delta/\lambda. \tag{3.9}$$

In context with on-off intermittency, the intermission is also called the off-time $\tau_{off}$ and the lifetime of a cluster the on-time $\tau_{on}$; both expressions will be used synonymously. Hence, the mean on- and off-times are defined by

$$\tau_{on} \equiv \tau_c = \overline{N_c}\,\lambda \quad \text{and} \quad \tau_{off} \equiv \delta \tag{3.10}$$

Table. 1. The statistical features of the Intermittent Poisson Process.

|  | Statistical variable | on-off intermittency | Characteristic function |
|---|---|---|---|
| inter-event time | $\lambda$ |  | $U_\lambda = 1/(1 - i2\pi f\lambda)$ |
| intermission = off-time | $\delta$ | $\tau_{off}$ | $U_\delta = 1/(1 - ir2\pi f\lambda)$ |
| lifetime of a cluster = on-time | $\tau_c = \overline{N_c}\,\lambda$ | $\tau_{on}$ | $U_{\tau_c} = \sum_{n=1}^{N_{max}} q_n U_\lambda^n$ |
| time between cluster heads | $\Lambda = \delta + \tau_c$ | $\Lambda = \tau_{off} + \tau_{on}$ | $U_\Lambda = U_\delta U_{\tau_c}$ |

### 3.3. *The diagonal and off-diagonal spectral contributions of the intermittent shot noise*

Based on the application of the Wiener-Khinchin theorem [2-4], the one-sided power spectrum of the intermittent shot noise is defined by

$$S_{shot}^{im}(f) = 2 \lim_{T\to\infty} \frac{\overline{Y(f,T)Y^*(f,T)}}{T}. \tag{3.11}$$

The star * indicates the conjugate complex. For convenience, we define

$$H_{j,\mu} \equiv H_{j,\mu}(f) \quad \text{and} \quad U_X \equiv U_X(f). \tag{3.12}$$

Applying (3.4) yields

$$\overline{Y(f,T)Y^*(f,T)} = \overline{\left[\sum_{j=1}^{N_c^T}\sum_{\mu=1}^{N_j} H_{j,\mu}\,exp(-2\pi f\Theta_{j,\mu})\right]\left[\sum_{j'=1}^{N_c^T}\sum_{\nu=1}^{N_{j'}} H^*_{j',\nu}\,exp(+i2\pi f\Theta_{j',\nu})\right]}. \tag{3.13}$$

The diagonal ($j = j'$) and off-diagonal terms ($j \neq j'$) are derived separately. Correspondingly, we define

$$S_{shot}^{im}(f) = S_{j=j'}(f) + S_{j\neq j'}(f) \tag{3.14}$$

whereby

$$S_{j=j'}(f) = 2\lim_{T\to\infty} \frac{\overline{Y(f,T)Y^*(f,T)}_{j=j'}}{T} \quad \text{and} \quad S_{j\neq j'}(f) = 2\lim_{T\to\infty} \frac{\overline{Y(f,T)Y^*(f,T)}_{j\neq j'}}{T} \tag{3.15}$$

is the power spectrum of the diagonal and off-diagonal contribution respectively. The corresponding derivations can be found in Appendices A, B and C.

### 4. The Power Spectrum of the Intermittent Shot Noise

According to Appendix (C.1), the power spectrum of the intermittent shot noise is

$$S_{shot}^{im}(f) = \frac{2}{\delta+\tau_c}\overline{N_c}\,\overline{|H(f)|^2} + \frac{2}{\delta+\tau_c}\left|\overline{H(f)}\right|^2 \Phi_{ex}(f). \tag{4.1}$$

For normal shot noise the rate of events is $1/\lambda$ (Fig. 2.a). For intermittent shot noise, the rate of events is reduced by the factor (Fig. 2.c)

$$\beta_{im} = \frac{\tau_c}{\delta+\tau_c} = \frac{\overline{N_c}}{r+\overline{N_c}}. \tag{4.2}$$

Applying this and (2.5), the first term on the right-hand side of (4.1) is identified as reduced shot noise

$$\beta_{im}\frac{2}{\lambda}\overline{|H(f)|^2} = \beta_{im}S_{shot}(f). \tag{4.3}$$

The second term in (4.1) is the spectral contribution of the excess noise; using (2.5), this can be expressed by

$$S_{ex}(f) = \frac{1}{2}S_{shot}(f)\frac{1}{r+\overline{N_c}}\Phi_{ex}(f). \tag{4.4}$$

Herein (Appendix (C.2))

$$\Phi_{ex}(f) = 2Re\left\{\frac{(1-U_\delta)(1-U_{\tau_c})}{U_\delta U_{\tau_c}-1}\frac{U_\lambda}{(U_\lambda-1)^2}\right\} \tag{4.5}$$

is the excess noise term; it is conserved under an exchange of $\delta$ and $\tau_c$. For $f \to 0$, the excess spectral function converges to

$$\Phi_{ex}(0) = \left(\frac{\delta}{\delta+\tau_c}\right)^2 \overline{N_c^2} - \left\{1 - \left(\frac{\tau_c}{\delta+\tau_c}\right)^2\right\}\overline{N_c}. \tag{4.6}$$

Using (4.3) and (4.4), the power spectrum of the intermittent shot noise can be expressed by

$$S^{im}_{shot}(f) = \beta_{im} S_{shot}(f) + \frac{1}{2} S_{shot}(f) \frac{1}{r+\overline{N_c}} \Phi_{ex}(f) \tag{4.7}$$

revealing a close relation between shot noise and excess noise.

## 5. Intermittent Shot Noise generating 1/f Fluctuations

In this Section, the intermittent shot noise is investigated for power-law distributed on-times. Correspondingly, the probability of finding exactly $n$ events in a cluster is described by a power-law like[2]

$$n = 1, 2, \ldots N_{max}: \qquad q_n = n^z / \sum_{n=1}^{N_{max}} n^z = q_1 \cdot n^z \tag{5.1}$$

applying for $-\infty < z < +\infty$; $N_{max}$ is a maximum cluster size. Applying (3.7), we obtain for

$$z = -2: \qquad \overline{N_c} \approx \frac{6}{\pi^2} \{\ln N_{max} + C_E\} \quad \text{and} \quad \overline{N_c^2} \approx \frac{6}{\pi^2} N_{max}. \tag{5.2}$$

$C_E = 0.5772\ldots$ is Euler's constant. Replacing the summation in (3.7) by integration, and conforming the results to (5.2), we find

$$\overline{N_c} \approx \frac{6}{\pi^2} \left\{ \frac{z+1}{z+2} \frac{N_{max}^{z+2}-1}{N_{max}^{z+1}-1} + C_E \right\} \quad \text{and} \quad \overline{N_c^2} \approx \frac{6}{\pi^2} \left\{ \frac{z+1}{z+3} \frac{N_{max}^{z+3}-1}{N_{max}^{z+1}-1} \right\}. \tag{5.3}$$

For large $N_{max}$, this converges to

$$-2 < z < -1: \qquad \log \overline{N_c} \approx (z+2) \log N_{max} \tag{5.4}$$

and

$$-3 < z < -1: \qquad \log \overline{N_c^2} \approx (z+3) \log N_{max}. \tag{5.5}$$

Fig. 4 shows the normalized logarithms of $\overline{N_c}$ and $\overline{N_c^2}$ as a function of exponent $z$.

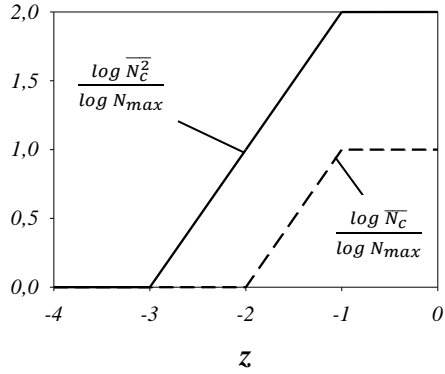

Fig. 4. The normalized logarithms of $\overline{N_c}$ (dashed) and of $\overline{N_c^2}$ (bold) for large $N_{max}$ as a function of exponent $z$.

### 5.1. *The spectral features of the excess noise*

Applying (5.1), the excess spectral function has been calculated on computer; it exhibits a 1/f shape which is denoted by $\Phi_{1/f}(f)$. Fig. 5 shows $\Phi_{1/f}(f)$ for $z = -2$ and $N_{max} = 10^6$ and for several values of normalized off-time $r$. The slope $b$ attains its maximum value for $r \gg 1$ (straight line); decreasing $r$, the slope $b$ declines until $r \approx 1$ (long-dashed line). A significant limit for the behavior of the slope is marked off by $r = 1$: For $r \leq 1$, the slope $b$ remains constant independent of $r$ and $\Phi_{1/f}(f) \propto r^2$ [20].

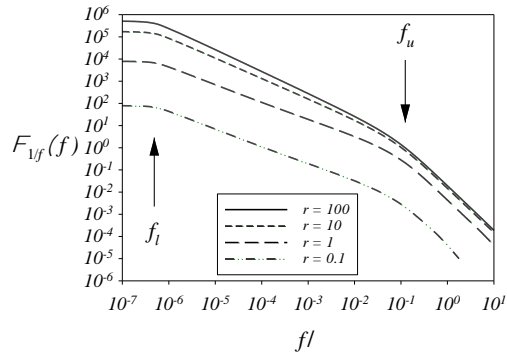

Fig. 5. The excess spectral function $\Phi_{1/f}(f)$ versus reduced frequency $f\lambda$ in a double-logarithmic scale for exponent $z = -2$ and for several values of normalized off-time $r$. The scaling region $N_{max} = 10^6$. $f_l$ and $f_u$ is the lower and upper cut-off frequency of 1/f noise, respectively.

---

[2] Likewise, the distribution of the on-times can be defined by $q_n \propto n^{-\mu_{on}}$; the exponents are related by $z = -\mu_{on}$.

## 5.2. Approximation for the 1/f^b shape within the scaling region

Within the scaling region, 1/f noise is excellently approximated by

$$\Phi_{1/f}(f) \approx \frac{C}{(f\lambda)^b} \tag{5.6}$$

scaling within the lower and upper cut-off frequency

$$f_l \approx 1/2N_{max}\lambda \quad \text{and} \quad f_u \approx 1/2\pi\lambda. \tag{5.7}$$

Below $f_l$, $\Phi_{1/f}(f)$ attains constant values; above $f_u$ it falls with $f^{-2}$. The scaling region is

$$f_u/f_l \approx N_{max} \tag{5.8}$$

is supposed to extend over many decades. In a logarithmic scale, $f_m$ is located in the middle of the scaling region (Fig. 6); $f_m$ is the geometric mean of $f_l$ and $f_u$ yielding

$$f_m = \sqrt{f_l f_u} \approx \frac{1}{\lambda\sqrt{4\pi N_{max}}}. \tag{5.9}$$

The slope $b$ and the amplitude $C$ in (5.6) are determined at $f_m$.

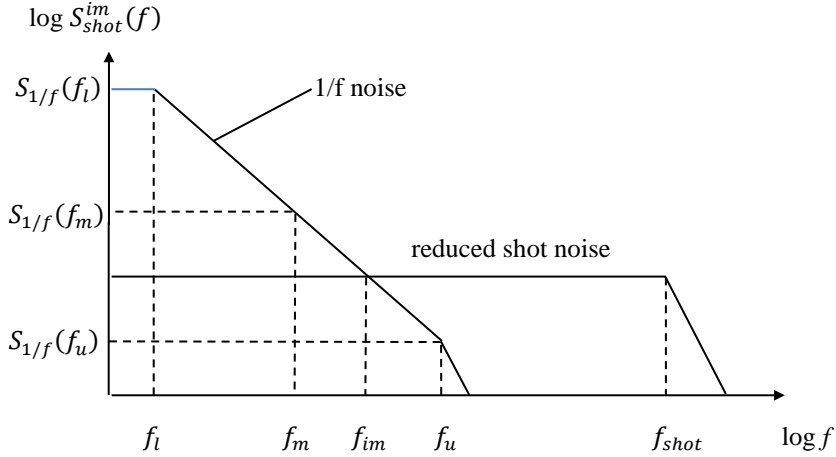

Fig. 6. The power spectrum of reduced shot noise and of 1/f noise with corresponding limiting frequencies.

For the following, we assume that the lifetimes of pulses are much smaller than the inter-event times ($\tau_h \ll \lambda$). Under this condition, the upper limit of 1/f noise $f_u \ll f_{shot} = 1/2\pi\tau_h$ leading to a well-extended shot noise plateau; within the scaling region $S_{shot}(f) \approx S_{shot}(0)$ (Fig. 6). Applying (4.4), the power spectrum of 1/f noise is obtained by

$$S_{1/f}(f) = \frac{1}{2} S_{shot}(0) \frac{1}{r+\overline{N_c}} \Phi_{1/f}(f). \tag{5.10}$$

The low frequency region is dominated by 1/f noise and the high frequency region by reduced shot noise (Fig. 6). $\Phi_{1/f}(f_m)$ being proportional to $S_{1/f}(f_m)$ is determined by

$$\Phi_{1/f}(f_m) = \sqrt{\Phi_{1/f}(f_l)\,\Phi_{1/f}(f_u)}. \tag{5.11}$$

As is seen in Fig. 6, $\Phi_{1/f}(f_l) \approx \Phi_{1/f}(0)$ which according to (4.6) yields[3]

$$\Phi_{1/f}(f_l) \approx \left(\frac{r}{r+\overline{N_c}}\right)^2 \overline{N_c^2}. \tag{5.12}$$

$\Phi_{1/f}(f_u)$ depends on the normalized off-time $r$ (Fig. 5). The results are summarized in Table 2.

Table. 2. The excess noise function $\Phi_{1/f}(f)$ at the lower and upper cut-off frequency and at the intermediate frequency $f_m$.

|  | $\Phi_{1/f}(f_l) \approx$ | $\Phi_{1/f}(f_u) \approx$ | $\Phi_{1/f}(f_m) \approx$ |
|---|---|---|---|
| $r \leq 1$ | $\left(\frac{r}{r+\overline{N_c}}\right)^2 \overline{N_c^2}$ | $r^2$ | $\frac{r^2}{r+\overline{N_c}}\sqrt{\overline{N_c^2}}$ |
| $r \geq 1$ |  | $1$ | $\frac{r}{r+\overline{N_c}}\sqrt{\overline{N_c^2}}$ |

---

[3] As is seen in Fig. 4, $\overline{N_c^2} \gg \overline{N_c}$ for all values of exponent $z$; this justifies neglecting the second term on the right-hand side of (4.6).

### 5.3. *The behavior of the slope b*

The slope $b$ in (5.6) is defined by (Fig. 6)

$$b = -\frac{\log S_{1/f}(f_l) - \log S_{1/f}(f_u)}{\log f_l - \log f_u}. \tag{5.13}$$

Applying (5.8) and (5.10), this is transformed into

$$b = \frac{\log\{\Phi_{1/f}(f_l)/\Phi_{1/f}(f_u)\}}{\log N_{max}}. \tag{5.14}$$

Using Table 2, the slope $b$ is provided for several values of normalized intermission $r$ (second column in Table 3). Applying Eqs. (5.4) and (5.5), the slope $b$ as a function of exponent $z$ is represented by the straight lines $b_1$ to $b_4$ (Fig. 7).

Table 3. Slopes $b$ and $\hat{b}$ and exponent $\hat{z}$ for several values of normalized intermission $r$.

| | | | |
|---|---|---|---|
| $r \lesssim 1$ | $b \approx \frac{\log \overline{N_c^2} - 2\cdot\log(\overline{N_c})}{\log N_{max}}$ | $\hat{b} \approx 1 - \frac{\log\left[\frac{6}{\pi^2}(\ln N_{max} + C_E)^2\right]}{\log N_{max}}$ | $\hat{z} = -2$ |
| $1 \leq r \leq N_{max}$ | $b \approx \frac{\log \overline{N_c^2} + 2\cdot\log r - 2\cdot\log(r+\overline{N_c})}{\log N_{max}}$ | $\hat{b} \approx 1 + \frac{\log r}{\log N_{max}}$ | $\hat{z} \approx -2 + \frac{\log r}{\log N_{max}}$ |
| $r \geq N_{max}$ | $b \approx \frac{\log \overline{N_c^2}}{\log N_{max}}$ | $\hat{b} = 2$ | $-1 \leq z$ |

As is seen in Fig. 7, the slope $b$ exhibits a distinct maximum; it is denoted by $\hat{b}$ and the corresponding exponent by $\hat{z}$. $\hat{b}$ is determined by applying (5.2), (5.4) and (5.5) (third column in Table 3). $\hat{z}$ is provided by the intersection of the straight lines $b_2$ and $b_3$ (fourth column in Table 3).

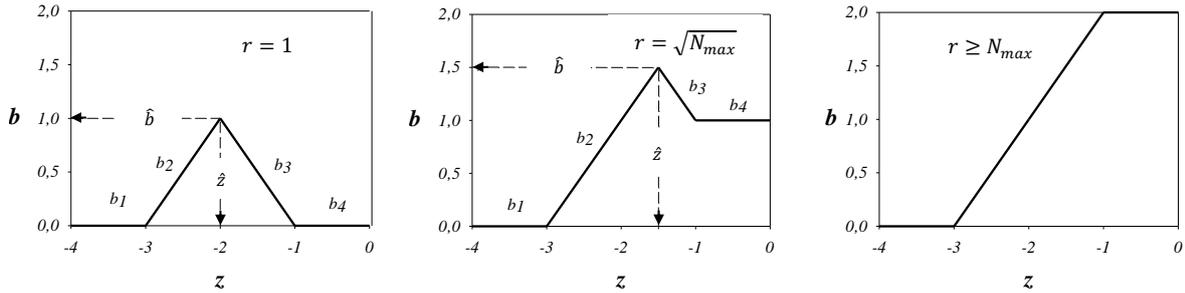

Fig. 7. The slope $b$ as a function of exponent $z$ for several values of normalized off-time $r$. The maximum value of the slope is denoted by $\hat{b}$ and the corresponding exponent by $\hat{z}$.

## 6. Preference for a pure 1/f shape

According to Table 3, for

$$1 \leq r \leq N_{max}: \qquad \hat{z} \approx -2 + \frac{\log r}{\log N_{max}} \quad \text{and} \quad \hat{b} \approx 1 + \frac{\log r}{\log N_{max}} \tag{6.1}$$

leading to $-2 \leq \hat{z} \leq -1$ and $1 \leq \hat{b} \leq 2$. For a scaling region $N_{max} \gg r$, these expressions converge to

$$\hat{z} \to -2 \quad \text{and} \quad \hat{b} \to 1 \tag{6.2}$$

exhibiting a pure 1/f shape. Such a preference for a pure 1/f shape is observed in many systems [7]. This justifies confining our further considerations to a slope $b = 1$. Under this condition, Eq. (5.6) reads

$$\Phi_{1/f}(f) \approx \frac{C}{f\lambda}. \tag{6.3}$$

Applying (4.7), the intermittent shot noise generating 1/f noise can be expressed by

$$S_{shot}^{im}(f) = S_{shot}(0)\left\{\frac{\beta_{im}}{1+(2\pi f\tau_h)^2} + \frac{1}{4}\frac{\alpha_{im}}{f\lambda}\right\}. \tag{6.4}$$

Herein the coefficient

$$\alpha_{im} = \frac{2C}{r+\overline{N_c}} \tag{6.5}$$

depends on the parameters of on-off intermittency. For the rectangular pulses seen in Fig. 1, the 1/f noise term in (6.4) can be expressed by

$$S_{1/f}(f) = \left(\frac{A\tau_h}{\lambda}\right)^2 \frac{\alpha_{im}}{f}. \tag{6.6}$$

The frequency where the 1/f shape is equal to reduced shot noise is found at (Fig. 6)

$$f_{im} = \frac{1}{4}\frac{\alpha_{im}}{\beta_{im}\lambda} = \frac{C}{2\overline{N_c}\lambda}. \tag{6.7}$$

For an increasing scaling region, $f_{im}$ is shifted to lower frequencies.

### 6.1. *The determination of the amplitude C*

The amplitude $C$ of 1/f noise in (6.3) is defined at $f_m$ yielding

$$C = f_m\lambda \cdot \Phi_{1/f}(f_m). \tag{6.8}$$

Applying (5.9) and the corresponding expressions in Table 2 we obtain the amplitude $C$ and the coefficient $\alpha_{im}$; the results are summarized in Table 4.

Table 4. The amplitude $C$ and of the coefficient $\alpha_{im}$ for several values of normalized intermission $r$.

|  | $f_m\lambda \approx$ | $\Phi_{1/f}(f_m) \approx$ | $C \approx$ | $\alpha_{im} =$ | $\alpha_{im} =$ |
|---|---|---|---|---|---|
| $r \leq 1$ | $\frac{1}{\sqrt{4\pi N_{max}}}$ | $\frac{r^2}{r+\overline{N_c}}\sqrt{0.6\,N_{max}}$ | $0.2\frac{r^2}{r+\overline{N_c}}$ | $0.4\left(\frac{r}{r+\overline{N_c}}\right)^2$ | $0.4\,(1-\beta_{im})^2$ |
| $r \geq 1$ |  | $\frac{r}{r+\overline{N_c}}\sqrt{0.6\,N_{max}}$ | $0.2\frac{r}{r+\overline{N_c}}$ | $0.4\frac{r}{(r+\overline{N_c})^2}$ | $0.4\,\frac{1-\beta_{im}}{r+\overline{N_c}}$ |

Considering

$$\frac{r}{r+\overline{N_c}} = 1 - \beta_{im} \tag{6.9}$$

the coefficient $\alpha_{im}$ can be expressed in terms of the coefficient $\beta_{im}$ (far right of Table 4). This allows comparing the spectral contribution of reduced shot noise with that of 1/f noise (see Eq. (6.4)).

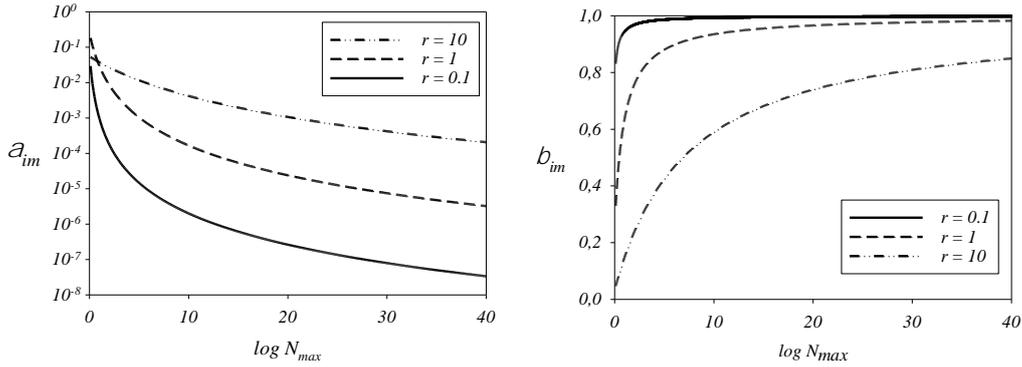

Fig. 8. The coefficients $\alpha_{im}$ and $\beta_{im}$ versus the logarithm of the scaling region $N_{max}$ for several values of normalized off-time $r$.

Fig. 8 shows the coefficients $\alpha_{im}$ and $\beta_{im}$ as a function of the scaling region $N_{max}$ for several values of the normalized off-state $r$. For a sufficiently large scaling region, the coefficient $\beta_{im}$ converges to 1 whereas the coefficient $\alpha_{im}$ approaches 0; in other words: as the full shot noise level is reached, the 1/f noise component disappears.

## 7. The Extreme Property of the Variance-Time Curve in the Presence of 1/f Fluctuations

It is well established that the variance of a 1/f noise process is much larger than the variance of a white noise process [21-22]. In this Section, we investigate the variance of the intermittent shot noise in the presence of 1/f fluctuations. As is seen in Fig. 2.c, the stochastic point process underlying intermittent shot noise is the Intermittent Poisson Process (= IPP) denoted by $y_{IPP}(t)$; this is described by a counting function

$$N_{IPP}(t) = \int_{t_0}^{t_0+t} y_{IPP}(t')dt' \tag{7.1}$$

representing the number of events that have occurred during a time from $t_0$ to $t_0 + t$. The variance of counts in a time interval $t$ is $V_{IPP}(t) \equiv var\{N_{IPP}(t)\}$ and is denoted the variance-time curve. Stochastic point processes generating 1/f noise are characterized by self-affinity [23]; scaling of time axis results in a statistical sense in an amplitude-scaled version of the same signal $N_{IPP}(at) \propto a^H N_{IPP}(t)$ applying within the scaling region $\tau_{min} < t < \tau_{max}$; $H$ is the so-called Hurst exponent [23]. This implies that for

$$\tau_{min} < t < \tau_{max}: \qquad V_{IPP}(t) \propto t^{2H}. \tag{7.2}$$

$\tau_{min}$ and $\tau_{max}$ correspond to the upper and lower frequency limit $f_u$ and $f_l$ respectively. For a well-extended scaling region $2H = 1 + b$ [23].

## 7.1. *The variance-time curve outside the scaling region*

Outside the scaling region, the following relations hold for the asymptotic behavior of the variance-time curve at short and long times respectively [24]

$t \leq \tau_{min}$:
$$V_0(t) = G_{IPP}(\infty)\, t \tag{7.3}$$

and

$t \geq \tau_{max}$:
$$V_\infty(t) = G_{IPP}(0)\, t. \tag{7.4}$$

$G_{IPP}(f)$ is the two-sided spectrum of the Intermittent Poisson Process (Fig. 2.c). $G_{IPP}(f)$ derives from the one-sided power spectrum in (4.1) by putting $\overline{|H(f)|^2} = \left|\overline{H(f)}\right|^2 = 1$ and dividing by 2; this yields

$$G_{IPP}(f) = \frac{1}{\delta + \tau_c}\{\overline{N_c} + \Phi_{ex}(f)\}. \tag{7.5}$$

This provides

$t \leq \tau_{min}$:
$$V_0(t) \equiv \frac{\overline{N_c}}{\delta + \tau_c} t = \frac{t}{\lambda}\beta_{im} \tag{7.6}$$

corresponding to reduced shot noise and in combination with (4.6)

$t \geq \tau_{max}$:
$$V_\infty(t) = \frac{t}{\lambda} C_V. \tag{7.7}$$

Herein

$$C_V \equiv \lambda\, G_{IPP}(0) = \left\{\frac{r^2\overline{N_c^2} + \overline{N_c}^3}{(r + \overline{N_c})^3}\right\} \tag{7.8}$$

is the pre-factor of $V_\infty(t)$ indicating the variance of counts in excess to $V_0(t)$ the variance of counts due to reduced shot noise. In the presence of 1/f fluctuations, $C_V$ depends on the exponent $z$ and on the normalized off-time $r$. Fig. 9 illustrates the variance-time curve in a double-logarithmic scale. Applying (5.3), the pre-factor $C_V$ has been calculated on computer. Fig. 10 shows $C_V$ as a function of the exponent $z$ for several values of the normalized off-time $r$.

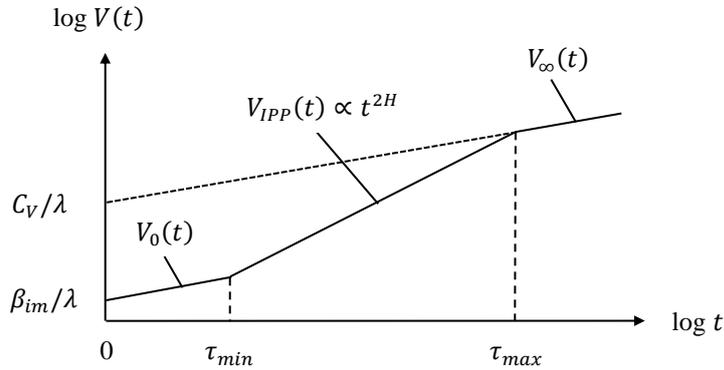

Fig. 9. Illustration of the variance-time curve in the presence of 1/f fluctuations.

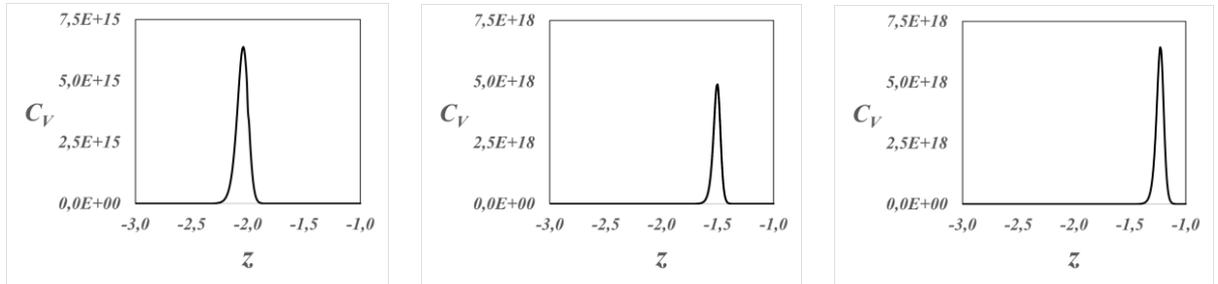

Fig. 10. The pre-factor $C_V$ as a function of exponent $z$ for several values of the normalized off-time $r$. Left: $r = 1$, middle: $r = 10^{10}$ and right: $r = 10^{15}$. The scaling region $N_{max} = 10^{20}$.

The pre-factor $C_V$ exhibits a distinct maximum; it is denoted $C_V^{max}$ and is located at $z_V$. In a logarithmic scale, $C_V$ is described by two straight lines. The exponent $z_V$ is determined by the intersection of these two straight lines yielding

$1 \leq r \leq N_{max}$:
$$z_V \approx -2 + \frac{\log r}{\log N_{max}} \qquad (7.9)$$

leading to $-2 < z_V < -1$. By comparison with (6.1) it is seen that

$$z_V = \hat{z} \qquad (7.10)$$

coinciding with the maximum value of the slope $b$. In summary it can be said, therefore, that the variance of counts attains extreme values which are accompanied by the extreme property of the slope $b$.

## 8. Results and Discussions

Shot noise is described by a random succession of elementary events (Fig. 1). Introducing a gating function, we obtain intermittent shot noise described by alternately succeeding intermissions and fluctuating clusters (Fig. 2). The power spectrum of intermittent shot noise is derived by applying the Wiener-Khinchin theorem. The derivations lead to a diagonal term comprising mutual time relations between events within a cluster and off-diagonal terms involving mutual time relations between events in different clusters. As a result, we obtain reduced shot and excess noise which can be expressed in terms of the characteristic functions of the on-off times.

We regard a power-law distributed cluster size being characterized by an exponent $z$. This leads to an excess noise which is excellently approximated by a $1/(f\lambda)^b$ shape; $\lambda$ is the inter-event time. The behavior of the slope $b$ depends (on a linear scale) on the exponent $z$ and (on a logarithmic scale) on the scaling region of 1/f noise and on the normalized off-time $r = \delta/\lambda$. If the off-time is shorter than the inter-event times ($\delta < \lambda$), the slope attains a maximum value $b \approx 1$ which is found for $z \approx -2$; if $\delta > \lambda$, this maximum value is found for $-2 < z < -1$ leading to $1 < b < 2$. For a scaling region much larger than the normalized off-time, we find a preference for a pure 1/f shape.

Finally, we investigate the variance of counts occurring within a time interval. In the presence of 1/f fluctuations, the variance of counts exhibits a maximum value which strongly depends on the normalized off-time $r$. The extreme property of variance of counts is accompanied by the extreme property of slope $b$.

Kononovicius and Kaulakys [25] derived the power spectrum of non-overlapping rectangular pulses separated by gaps; they investigated the case of power-law distributed gaps generating 1/f noise. This can be compared with the intermittent shot noise generating 1/f noise where off-times (= gaps) are exponentially, and on-times (= rectangular pulses) are power-law distributed. Since the power spectrum of the intermittent shot noise is conserved under an exchange of on- and off-times, the results of Kononovicius and Kaulakys correspond to ours.

The above-mentioned controversial debate on 1/f noise in semiconductors has been enriched by the application of the intermittent-shot-noise model. Traps in the semiconductor material are thought to generate charge carriers intermittently [26]; this leads to fluctuating clusters that result in a 1/f shape. A physical interpretation of such a behavior is still missing.

The intermittent-shot-noise model may find further applications where intermittently occurring spikes or pulses are accompanied by 1/f fluctuations. For example, this may be the case for neuronal spike trains observed in cortical neurons [27]; the parameters of the intermittent process adapted to empirically observed 1/f spectra could characterize the activity of different neurons. A further possible application concerns excess noise of dc current-carrying thin metal wires in dilute gases [28]. The origin of excess noise – exhibiting a pure 1/f shape – is supposed to be in the boundary between the metal and gas; adsorbed gas atoms may lead to an intermittent current and heat flow.

**Declaration of competing interest**

The author declares that they have no known competing financial interests or personal relationships that could have influenced the work reported in this paper.

**Acknowledgment**

The author would like to thank Laszlo Kish for inspiring discussions and Barbara Herzberger for proofreading this manuscript.

## Appendix A. The Spectral Contribution of the Diagonal Term

Using (3.13), the spectral contribution of the diagonal term can be written as

$$\overline{Y(f,T)Y^*(f,T)}_{j=j\prime} = \overline{\sum_{j=1}^{N_c^T}\left[\sum_{\mu=1}^{N_j}|H_{j,\mu}|^2 + 2Re\left\{\sum_{\mu=1}^{N_j-1}\sum_{\nu=1}^{N_j-\mu} H_{j,\mu} H^*{}_{j,\mu+\nu} \exp[-i2\pi f(\Theta_{j,\mu} - \Theta_{j,\mu+\nu})]\right\}\right]}. \quad (A.1)$$

Number fluctuations are described by the cluster size distribution $q_n = prob\{N_j = n\}$ with $n = 1, 2, \ldots N_{max}$. By averaging over all possible Poisson fragments, the first term in square brackets yields

$$\overline{\sum_{j=1}^{N_c^T}\sum_{\mu=1}^{N_j}|H_{j,\mu}|^2} = N_c^T \sum_{n=1}^{N_{max}} q_n \sum_{\mu=1}^{n}\overline{|H|^2} = N_c^T \overline{N_c} \, \overline{|H|^2}. \quad (A.2)$$

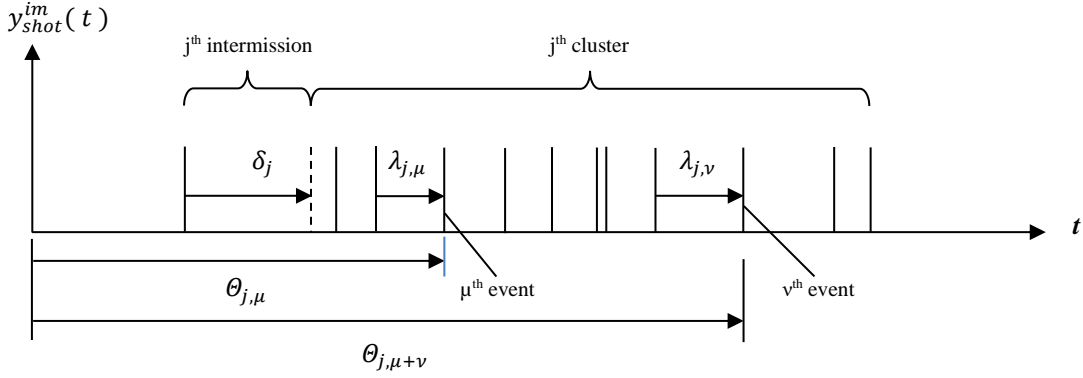

Fig. 12. Occurrence times of the μ$^{th}$ and ν$^{th}$ event within the j$^{th}$ cluster. Each spike trigger a pulse $h(t)$ (not shown in this Figure).

The time interval between the μ$^{th}$ and ν$^{th}$ event in the j$^{th}$ cluster can be expressed by (Fig. 12)

$$-(\Theta_{j,\mu} - \Theta_{j,\mu+\nu}) = \sum_{s=1}^{\nu} \lambda_{j,s} \tag{A.3}$$

Considering the assumptions made in Section 3.2, Eq. (A.1) can be written as

$$\overline{Y(f,T)Y^*(f,T)_{j=j\prime}} = N_c^T \left[ \overline{N_c} \, \overline{|H|^2} + \left|\overline{H}\right|^2 2Re\left\{ \sum_{n=1}^{N_{max}} q_n \sum_{\mu=1}^{n-1} \sum_{\nu=1}^{n-\mu} U_\lambda^\nu \right\} \right]. \tag{A.4}$$

Applying $\sum_{i=1}^{n} y^{i-1} = (1 - y^n)/(1 - y)$, the real part yields

$$\overline{N_c} \, 2Re\left\{ \frac{U_\lambda}{1-U_\lambda} \right\} + \Phi_c(f) \tag{A.5}$$

where

$$\Phi_c(f) \equiv 2Re\left\{ (U_{\tau_c} - 1) \frac{U_\lambda}{(U_\lambda - 1)^2} \right\} \tag{A.6}$$

comprising mutual time correlations among events within a cluster. $N_c^T$ in (A.4) is the only term depending on the time interval $T$. Considering that the average time between cluster heads is $\delta + \tau_c$ and that there are exactly $N_c^T$ clusters within time interval $T$ yields

$$\lim_{T \to \infty} \frac{N_c^T}{T} = \frac{1}{\delta + \tau_c}. \tag{A.7}$$

Substituting (A.4) into the left-hand side of (3.16), the power spectrum of the diagonal term is given by

$$S_{j=j\prime}(f) = \frac{2}{\delta + \tau_c} \left\{ \overline{N_c} \, \overline{|H(f)|^2} + \overline{N_c} \, \left|\overline{H(f)}\right|^2 2Re\left( \frac{U_\lambda}{1-U_\lambda} \right) + \left|\overline{H(f)}\right|^2 \Phi_c(f) \right\} \tag{A.8}$$

applying for arbitrarily distributed intermissions $\delta$, inter-event times $\lambda$ and cluster size distributions $q_n$. For exponentially distributed inter-event times defined in Section (3.2), $Re\{U_\lambda/(1 - U_\lambda)\} = 0$ reducing (A.8) to

$$S_{j=j\prime}(f) = \frac{2}{\delta + \tau_c} \left\{ \overline{N_c} \, \overline{|H(f)|^2} + \left|\overline{H(f)}\right|^2 \Phi_c(f) \right\}. \tag{A.9}$$

**Appendix B. The Spectral Contribution of the Off-diagonal Term**

Replacing j' by j+k and rearranging the terms in the double-sum of (3.13), the off-diagonal term can be written as

$$\overline{Y(f,T)Y^*(f,T)_{j \neq j\prime}} = 2Re\left\{ \sum_{j=1}^{N_c^T - 1} \sum_{k=1}^{N_c^T - j} \sum_{\mu=1}^{N_j} \sum_{\nu=1}^{N_{j+k}} H_{j,\mu} H^*_{j+k,\nu} exp[-i2\pi f(\Theta_{j,\mu} - \Theta_{j+k,\nu})] \right\}. \tag{B.1}$$

The time between the μ$^{th}$ event in the j$^{th}$ cluster and the ν$^{th}$ event in the (j+k)$^{th}$ cluster can be expressed by (Fig. 13)

$$-(\Theta_{j,\mu} - \Theta_{j+k,\nu}) = \tilde{\vartheta}_{j,\mu} + \sum_{i=j+1}^{j+k-1} \Lambda_i + \vartheta_{j+k,\nu}. \tag{B.2}$$

Herein the occurrence time of the μ$^{th}$ event in the j$^{th}$ cluster and of the ν$^{th}$ event in the (j+k)$^{th}$ cluster respectively is

$$\tilde{\vartheta}_{j,\mu} = \sum_{s=\mu+1}^{N_j} \lambda_{j,s} \quad \text{and} \quad \vartheta_{j+k,\nu} = \delta_{j+k} + \sum_{s=1}^{\nu} \lambda_{j+k,s}. \tag{B.3}$$

$$\Lambda_i = \delta_i + \sum_{m=1}^{N_i} \lambda_{i,m} \tag{B.5}$$

is the time between cluster heads in the i$^{th}$ cluster. Substituting this into (B.1) yields

$$\overline{Y(f,T)Y^*(f,T)_{j \neq j\prime}} = \left|\overline{H(f)}\right|^2 2Re\left\{ R_1(f) R_2(f) \overline{\sum_{j=1}^{N_c^T - 1} \sum_{k=1}^{N_c^T - j} U_\Lambda^{k-1}} \right\} \tag{B.6}$$

where

$$R_1(f) = \overline{\sum_{\mu=1}^{N_j} exp(i2\pi f \tilde{\vartheta}_{j,\mu})} = \overline{\sum_{\mu=1}^{N_j} U_\lambda^{\mu-1}} = \frac{U_{\tau_c} - 1}{U_\lambda - 1} \tag{B.7}$$

and

$$R_2(f) = \overline{\sum_{\nu=1}^{N_{j+k}} exp(i2\pi f \vartheta_{j+k,\nu})} = U_\delta U_\lambda \frac{U_{\tau_c}-1}{U_\lambda - 1}. \tag{B.8}$$

The double sum in (B.6) leads to

$$\sum_{j=1}^{N_c^T-1} \sum_{k=1}^{N_c^T-j} U_\Lambda^{k-1} = N_c^T \frac{1}{1-U_\Lambda} + \frac{U_\Lambda^{N_c^T}-1}{(U_\Lambda-1)^2}. \tag{B.9}$$

The second term on the right-hand side can be shown to provide a spectral line at $f = 0$; it is omitted for the following. Applying (A.7), substituting (B.9), $R_1(f)$ and $R_2(f)$ into (B.6) and making use of the right-hand side of (3.15), the power spectrum of the off-diagonal term is obtained by

$$S_{j \neq j'}(f) = \frac{2}{\delta + \tau_c} \left|\overline{H(f)}\right|^2 \Phi_{im}(f). \tag{B.10}$$

Herein

$$\Phi_{im}(f) \equiv 2Re\left\{\frac{R_1(f)\,R_2(f)}{1-U_\Lambda}\right\} = 2Re\left\{\frac{U_\delta(U_{\tau_c}-1)^2}{1-U_\delta U_{\tau_c}} \frac{U_\lambda}{(U_\lambda-1)^2}\right\} \tag{B.11}$$

is a spectral function describing mutual time correlations among events in different clusters. (B.11) applies for arbitrarily distributed intermissions $\delta$, inter-event times $\lambda$ and cluster size distributions $q_n$.

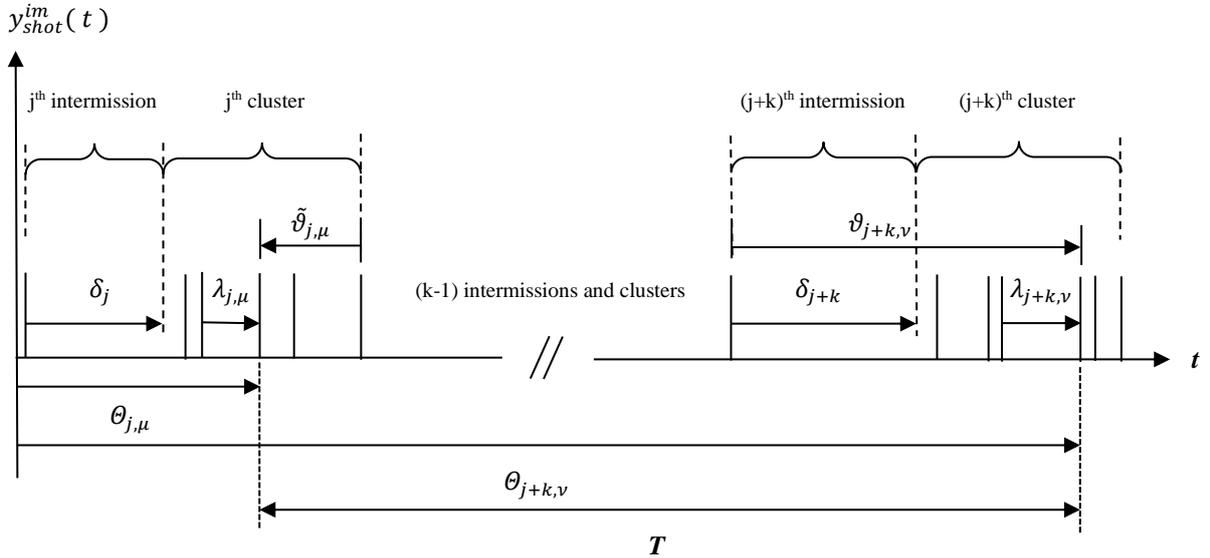

Fig. 13. The time interval $T$ starts at the $\mu^{th}$ event in the $j^{th}$ cluster and ends at the $\nu^{th}$ event in the $(j+k)^{th}$ cluster. There are (k-1) intermissions and clusters between the $j^{th}$ and the $(j+k)^{th}$ cluster (not illustrated in this Figure). Each spike trigger a pulse $h(t)$ (not shown in this Figure).

**Appendix C. The Power Spectrum of the Intermittent Shot Noise**

According to (3.15), (A.9) and (B.10), the power spectrum of the intermittent shot noise is obtained by

$$S_{shot}^{im}(f) = S_{j=j'}(f) + S_{j \neq j'}(f) = \frac{2}{\delta + \tau_c} \overline{N_c} \,\overline{|H(f)|^2} + \frac{2}{\delta + \tau_c} \left|\overline{H(f)}\right|^2 \Phi_{ex}(f). \tag{C.1}$$

Herein the excess noise term is

$$\Phi_{ex}(f) = \Phi_c(f) + \Phi_{im}(f) = 2Re\left\{\frac{(1-U_\delta)(1-U_{\tau_c})}{U_\delta U_{\tau_c}-1} \frac{U_\lambda}{(U_\lambda-1)^2}\right\}. \tag{C.2}$$